\documentclass[notitlepage,aps,pra,reprint,twocolumn,longbibliography,superscriptaddress]
{revtex4-1}
\usepackage{graphicx}
\usepackage{amsmath}
\usepackage{amssymb}
\usepackage{comment}
\usepackage[colorlinks, allcolors=blue]{hyperref}
\usepackage[all]{hypcap}
\usepackage[mathlines]{lineno}
\usepackage{physics}
\usepackage{wrapfig}
\usepackage{lipsum}
\usepackage{ulem}
\usepackage{siunitx}

\newcommand{\pref}[2]{\hyperref[#1]{\ref{#1}(#2)}}
\newcommand{\preff}[2]{\hyperref[#1]{\ref{#1}#2}}
\newcommand{\eqpref}[1]{\hyperref[#1]{(\ref{#1})}}

\newcommand{\squig}{{\raise.17ex\hbox{$\scriptstyle\sim$}}}

\begin{document}
\title{Spectroscopy of momentum state lattices}
\author{Sai Naga Manoj Paladugu}
\thanks{These authors contributed equally to this work.}
\affiliation{Department of Physics, University of Illinois Urbana-Champaign, Urbana, IL
61801-3080, USA}
\author{Tao Chen}
\thanks{These authors contributed equally to this work.}
\affiliation{Department of Physics, University of Illinois Urbana-Champaign, Urbana, IL
61801-3080, USA}
\affiliation{Interdisciplinary Center of Quantum Information, State Key Laboratory of
Modern Optical Instrumentation, and Zhejiang Province Key Laboratory of Quantum Technology
and Device of Physics Department, Zhejiang University, Hangzhou 310027, China}
\author{Fangzhao Alex An}
\thanks{These authors contributed equally to this work.}
\affiliation{Department of Physics, University of Illinois Urbana-Champaign, Urbana, IL
61801-3080, USA}
\author{Bo Yan}
\affiliation{Interdisciplinary Center of Quantum Information, State Key Laboratory of
Modern Optical Instrumentation, and Zhejiang Province Key Laboratory of Quantum Technology
and Device of Physics Department, Zhejiang University, Hangzhou 310027, China}
\author{Bryce Gadway}
\email{bgadway@illinois.edu}
\affiliation{Department of Physics, University of Illinois Urbana-Champaign, Urbana, IL
61801-3080, USA}
\date{\today}

\begin{abstract}
We explore a technique for probing energy spectra in synthetic lattices that is
analogous to scanning tunneling microscopy.
Using one-dimensional synthetic lattices of coupled atomic momentum states, we explore
this spectroscopic technique and observe qualitative
agreement between the measured and simulated energy spectra for small two- and
three-site lattices as well as a uniform many-site lattice. Finally, through
simulations, we show that this technique should allow for the exploration of the
topological bands and the fractal energy spectrum of the Hofstadter model as realized in
synthetic lattices.
\end{abstract}
\maketitle

\section{Introduction}
Cold atom systems are well suited to engineering Hamiltonians for the exploration of
condensed matter physics phenomena.
To help explore these systems, a suite of techniques have been developed over the past
decades to reveal the energy spectra in cold atom experiments, including injection 
spectroscopy based on auxiliary spin components~\cite{Gaebler2010,Cheuk-spin-spec,Zhang1},
lattice amplitude modulation spectroscopy~\cite{Greiner2002,Stoferle-Wiggle}, phasonic
modulation spectroscopy in quasiperiodic lattices~\cite{Rajagopal-spec-quasicrystal},
momentum-resolved band spectroscopy~\cite{Ernst2010,WeitSpec}, and Fourier transform
spectroscopy~\cite{Vald_s_Curiel_2017}. On their own, such probes are primarily global and do not provide
a direct window into the local spatial structure of states that comprise the spectrum
of a system.

Local spectroscopy would be of particular interest for certain problems related to
disordered~\cite{Sanchez-Palencia2010} or topological~\cite{CooperSpielmanRMP} systems.
For example, in topological systems, local spectroscopic probing at the boundary or within
the bulk could reveal distinct responses and serve to indicate the presence of topological
boundary modes. In disordered or quasiperiodic systems, the probing of mobility
edges~\cite{Semeghini2015,Luschen-ME,an2021interactions,Ganeshan,An-ME2} could be better facilitated
by local probes that distinguish between metallic and
insulating states in a energy-resolved manner.

We demonstrate such a local spectroscopic probe that is suitable for synthetic lattice
of coupled momentum states.
Similar spin-injection spectroscopy techniques have been used to explore spin-orbit coupling
in atomic Fermi gases~\cite{Gaebler2010,Cheuk-spin-spec,Zhang1}.
Indeed, it is natural to consider the extension of such techniques to 
synthetic lattices~\cite{Ozawa2019} that consist of discrete states~\cite{Sthul2015,Mancini2015,chalopin2020probing,roell2022chiral}, and in
particular, discrete momentum states~\cite{li2022abcaging,guo2020tunable,wang2022obs}.
By considering part of a synthetic lattice as a ``probe'' attached to a ``system'' of interest, and using the suite of controls afforded in synthetic lattice experiments, we study the energy-dependence of probe-system coupling to directly determine the energy spectrum of
dressed states in a synthetic lattice system.
This builds upon previous explorations using coupled momentum states that used system-reservoir coupling to engineer effective non-Hermitian loss~\cite{Lapp_2019,Tao-AB-ring}, as well as related demonstrations of energy-resolved spectroscopy in topological synthetic lattices of microwave-coupled Rydberg
levels~\cite{kanungo2022realizing} and non-Hermitian momentum state lattices~\cite{Liang2022}.
We demonstrate this technique on the simple test cases of few-site and many-site tight-binding lattices, finding qualitative agreement with theoretical predictions as well as observing the influence of atomic interactions. Using numerical simulations, we demonstrate the applicability of this technique for studies of topological band structures, including the celebrated Hofstadter butterfly spectrum and its associated topological boundary states.

This paper as organized as follows: In Sec.~\ref{theory}, we introduce the theoretical
underpinning behind our technique, based on Fermi's golden rule. In Sec.~\ref{methods}, we discuss the experimental methods used to perform injection spectroscopy on engineered momentum state lattices. In Sec.~\ref{expt}, we present the results of injection spectroscopy on exemplary systems that include simple two- and three-site lattice, as well as a uniform 26-site lattice. We compare our experimental results with
simulations based on numerical integration of the Gross-Pitaevskii equation (GPE). In Sec.~\ref{sims}, we use numerical simulations to demonstrate how this technique can be straightforwardly applied to observe the Hofstadter butterfly spectrum as well as topological bands in the 
Harper-Hofstadter model. Additionally, we describe extensions of the technique that allow for scanning-mode access to local spectroscopic signatures of extended systems. Finally, in Sec.~\ref{conc}, we conclude and discuss further prospects of this technique.

\begin{figure*}[t]
    \centering
    \includegraphics[width=1.9\columnwidth]{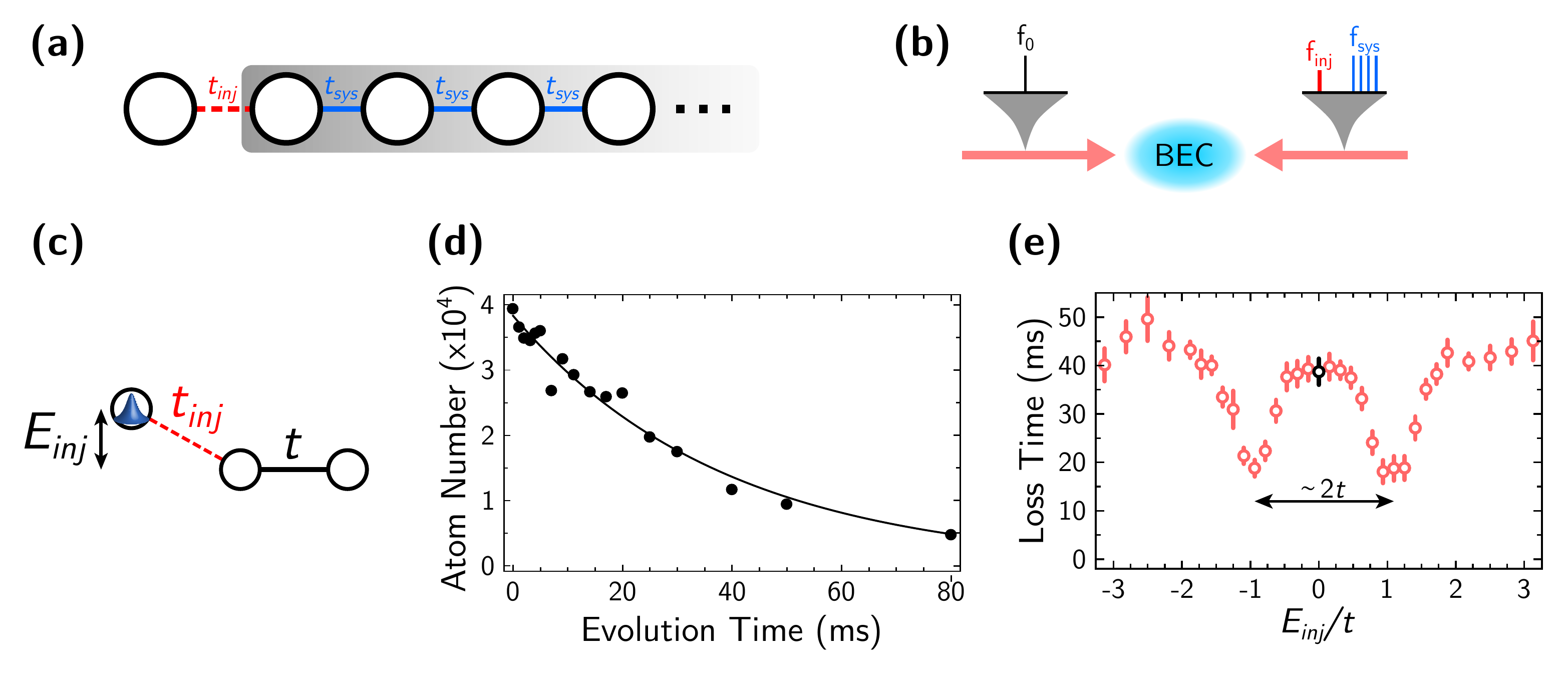}
    \caption{
    \textbf{Energy-resolved injection spectroscopy in synthetic momentum state lattices.}
    \textbf{(a)}~To probe a lattice system having characteristic tunneling energy $t_\textrm{sys}$, population is weakly injected in from a nearby probe site through a probe-system coupling term $t_\textrm{inj}$.
    \textbf{(b)}~In synthetic lattices of atomic momentum states, the system, probe, and all relevant coupling terms can be engineered through the spectral addressing of unique Bragg resonances. Here, two counter-propagating beams interfere to drive Bragg transitions between adjacent states with momenta $2 n \hbar k$ and $2 (n+1) \hbar k$, where $k = 2\pi/\lambda$ and $\lambda$ is the wavelength of the driving laser fields ($\SI{1064}{\nano\meter}$ in experiment).
    \textbf{(c)}~Illustration of the simplest two-site lattice system with inter-site hopping term $t  = h\times\SI{1598\pm 10}{\hertz}$, along with a  control of the energy bias $E_\textrm{inj}$ of the probe site relative to zero energy [$t_\textrm{inj} = h\times \SI{40\pm 10}{\hertz}$]. In practice, this bias tuning is accomplished simply through a change in the frequency of the applied injection tone, \textit{i.e.}, the spectral component labeled $f_\textrm{inj}$ in (b).
    In the weak-coupling limit ($t_\textrm{inj} \ll t_\textrm{sys})$, the rate of population loss from the probe into the system provides a measure of the local density of system states at the injection site.
    \textbf{(d)}~For injection into the two-site system at zero bias, we show the dynamics of the measured atomic population in the probe site, along with an exponential fit used to extract a characteristic loss time.
    \textbf{(e)}~Measured loss time plotted versus the injection site energy bias. The black data point at
    $E_\textrm{inj}/t = 0$ relates to the loss time extracted from (d). The two peaks of enhanced loss rates, centered around $E_\textrm{inj} = \pm t$, correspond to the symmetric
    and anti-symmetric eigenstates of the double well system.
    The error bars in (e) relate to the standard error of the fit-determined $1/e$ loss times.
    }
    \label{fig:scheme}
\end{figure*}

\section{Theory}
\label{theory}

We begin by considering the following Hamiltonian that describes the injection site, the discrete lattice under study, and their coupling:
\begin{align}
    \hat{H} &= \hat{H}_\textrm{inj} + \hat{H}_\textrm{latt} \\
    \hat{H}_\textrm{inj} &= t_\textrm{inj}\ket{0}\bra{1} + t_\textrm{inj}\ket{1}\bra{0}
    + E_\textrm{inj}\ket{0}\bra{0} \\
    \hat{H}_\textrm{latt} &= \sum\limits_{j=1}^{N-1} (t_{j}\ket{j+1}\bra{j} + \textrm{h.c.})
    + \sum\limits_{j=1}^N E_j\ket{j}\bra{j}.\label{eq:H}
\end{align}
Here, $\ket{0}$ represents the probe site, $t_\textrm{inj}$ is the injection tunneling between the
probe site and the lattice, $E_\textrm{inj}$ is the energy detuning of the
probe site with respect to the lattice, $N$ is the number of lattice sites, and the $t_j$
are the (in general link-specific) tunneling terms within the lattice itself. We restrict $t_\textrm{inj}$ to
be real-valued, while the $t_j$ can be complex. Let the $\epsilon_i$ be the eigenvalues and
$\ket{\psi_i}$ be the eigenvectors for the lattice Hamiltonian, Eq.~\ref{eq:H}. In general, we can write
\begin{align}
\ket{\psi_i} = \sum\limits_{n=1}^N c_n^{(i)}\ket{n}.
\end{align}
If $t_\textrm{inj}
\ll t_j$, then we can treat the injection Hamiltonian as a perturbation. In this limit, we
can use Fermi's golden rule in order to characterize the loss rate from the probe site into the lattice:
\begin{align}
    \Gamma &= \frac{2\pi}{\hbar}\sum\limits_{i=1}^N
    |\bra{\psi_i}\hat{H}_\textrm{inj}\ket{0}|^2
    \delta(E_i - E_\textrm{inj}) \\
    &= \frac{2\pi}{\hbar} t_\textrm{inj}^2 \sum\limits_{i=1}^N |c_1^{(i)}|^2
    \delta(E_i - E_\textrm{inj}).
\end{align}
In a real experiment, the delta function will become regularized
due to the finite amount of evolution time,
resulting in a Fourier-limited energy resolution.
Still, for sufficiently long evolution times and sufficiently small values of $t_\textrm{inj}$, a measurement of the loss rate as a function of $E_\textrm{inj}$ will permit an energy-resolved measurement of the local density of states.
Roughly speaking, the loss rate from the probe site will be enhanced if $E_\textrm{inj}$ is set close to the energy of a lattice eigenstates and will vanish if there are no lattice eigenstates in the vicinity of  $E_\textrm{inj}$.

One useful feature of injection spectroscopy is its sensitivity to the details of the eigenstate weight at the site of injection (\textit{i.e.}, its local nature). At the resonance condition ($E_\textrm{inj} = E_i$) for some eigenstate $\ket{\psi_i}$, the rate of loss from the probe site will be proportional to the overlap, or Franck-Condon factor, 
$|\braket{1}{\psi_i}|^2$. The more weight the eigenstate has with the site of injection, the
larger the loss rate. This sensitivity to the \textit{local} density of states should prove useful when, for example, probing the distinction between the bulk and edge spectra of a topological system ~\cite{goldman2013direct,Sthul2015,Mancini2015}.
In disordered or pseudo-disordered systems, this feature can also be useful for detecting metal-insulator transitions and for identifying mobility edges~\cite{an2018engine,an2021interactions}.

\begin{figure*}[t]
    \centering
    \includegraphics[width=1.9\columnwidth]{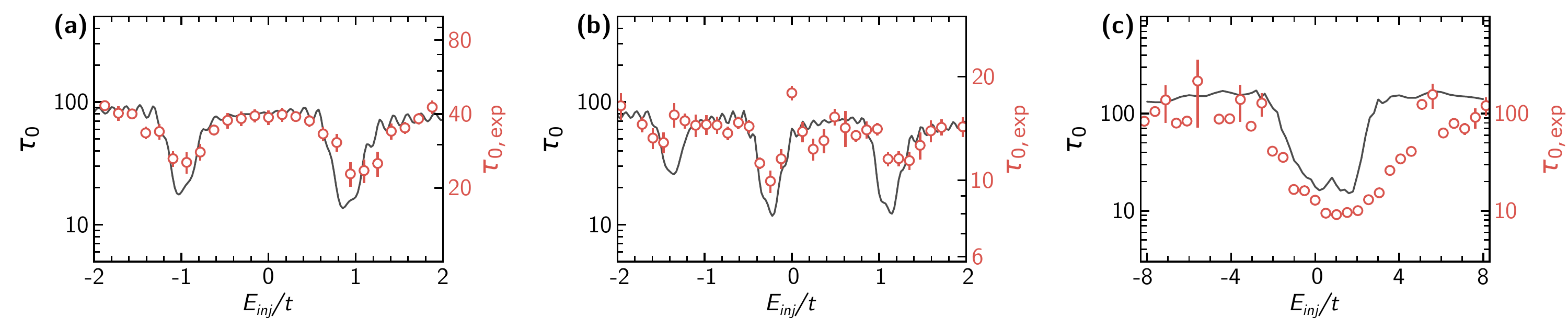}
    \caption{\textbf{Experimental and theoretical loss times for three different lattices.}
    The simulated curves are in black and the
    experimental data are in red.
    \textbf{(a)}~The loss times for
    spectroscopy of a two-site lattice
    as a function of $E_\textrm{inj}$ (in units of the lattice hopping energy $t$).
    This panel of data is the same as in 
    Fig.~\pref{fig:scheme}{a}, with $t = h\times\SI{1598\pm 10}{\hertz}$ and $t_\textrm{inj} = h\times\SI{40\pm 10}{\hertz}$. Note that the vertical axes for the simulation and the experimental data are on different scales.
    \textbf{(b)}~The loss times for injection into a three-site lattice with uniform system hopping $t = h\times\SI{2040\pm 3}{\hertz}$ and $t_\textrm{inj} = h\times\SI{50\pm 3}{\hertz}$. We again note the different vertical axes for the simulation and experimental data.
    \textbf{(c)}~The loss time as measured by injection spectroscopy of a uniform 26-site lattice.
    Here, we operate with a uniform system hopping $t = h\times\SI{492\pm 10}{\hertz}$ and with $t_\textrm{inj} = h\times \SI{50\pm 10}{\hertz}$, and the simulations and data have common vertical axes.
    The error bars in (a) and (b) reflect
    the standard error of the fit $1/e$ loss
    time. The error bars in (c) relate to
    the standard error of the measured
    probe population, propagated to an error
    in the $1/e$ time constant.
    }
    \label{fig:sims}
\end{figure*}

\section{Methods}
\label{methods}

In our experiment, we typically start with nearly pure
Bose--Einstein condensates (BEC) of roughly
$5\times10^4$ $^{87}\textrm{Rb}$
atoms confined in a crossed-dipole
trap formed from lasers of wavelengths $\SI{1064}{\nano\meter}$ and $\SI{1070}{\nano\meter}$. These trapping
beams create a harmonic trap with trap frequencies $\omega_{x,y,z} \approx 2\pi\times
\{130,10,130\}$~$\SI{}{\hertz}$. After forming the BEC, we suddenly turn on a retro-reflected path of the $\SI{1064}{\nano\meter}$ beam that contains a tailored frequency spectrum of discrete components. The resulting interference of the two counter-propagating beams having wavelength $\lambda = \SI{1064}{\nano\meter}$ serves to couple discrete atomic momentum orders $p_n = 2n\hbar k$ (with $k = 2\pi/\lambda$) via two-photon Bragg transitions. As summarized in Fig.~\ref{fig:scheme}, the Bragg transitions serve to form a synthetic lattice of momentum states~\cite{Gadway-KSPACE,Meier-AtomOptics}, as well as introduce an injection link between a probe site and the lattices under study.

%
%
More explicitly, the initially populated $p = 0$ momentum
order serves as the ``probe site,'' and resides next to the lattice to be probed.
We couple the probe site to an injection site of the lattice via a weak Bragg link, having an energy scale $t_\textrm{inj}$, as shown in 
Fig.~\pref{fig:scheme}{a}.
The probe site has a controllable energy bias relative to the lattice system, $E_\textrm{inj}$, which is introduced via a detuning of the probe-system Bragg transition, $\ket{0} \leftrightarrow \ket{1}$. 
The lattice system under study is composed by coupling (via Bragg transitions) the momentum orders $p = 2n\hbar k$, where $n \geq 1$ and is truncated at a final value depending on the size of the lattice under study. In the described experiments, the lattices we consider involve no variations of their potential landscape, and are thus formed by resonantly coupling all of the relevant momentum orders. In the Aubry-Andr\'e-Harper-Hofstadter model~\cite{Harper1955,hofstadter1976energy,aubry1980analyticity} considered in simulations, detunings of the Bragg transitions are used to introduce quasiperiodic variations in the site energies. 

In our experiments, all atoms initially start in the $p = 0$
order. For one experimental run, we fix $E_\textrm{inj}$ and let the atoms evolve
under the influence of our engineered Hamiltonian for a variable duration $\tau$. We then extract
the population of the $p = 0$ state versus the evolution time $\tau$, and fit this to an
exponential, $e^{-\tau/\tau_0}$. We repeat the experiment for many different
$E_\textrm{inj}$, and in the end we plot $\tau_0$, the loss time, as a function of
$E_\textrm{inj}$. When $E_\textrm{inj}$ is close in energy to an eigenstate of the
lattice, there will be a decrease in $\tau_0$. This reflects the strong enhancement of the transition rate from the probe site when $E_\textrm{inj}$ is in the vicinity of a lattice eigenenergy, due to the technique's sensitivity to the system density of states.

The ideal tight binding Hamiltonian (Eq.~\ref{eq:H}) does not account for off-resonant Bragg transitions and 
the inhomogeneous (in real-space) many-body interactions in the momentum state lattice. These introduce on-site energy shifts~\cite{an2018corr,Tao-AB-ring,An-Zhang} and consequently affect our experimental observations. 
To fully account for the effects of interactions, we perform 3D mean-field Gross-Pitaevskii equation (GPE) simulations taking into account the experimentally measured atom number, 
trap frequencies, and hopping amplitudes (Bragg field strengths). The details on how to resolve the momentum
space dynamics with the time evolution of spatial GPE are described in Ref.~\cite{Chen2021-gpe}. The loss time 
constant $\tau_0$ is calculated from the dynamics of the number of atoms remaining in the original condensate momentum order ($p=0$), normalized to the initial atom number. While the integration of the GPE under a time-dependent
Bragg field can help to address the effects from mean-field interactions and the external dipole trap, the 
decoherence caused by long-time thermal fluctuations (\textit{e.g.}, inelastic collisions between different 
momentum states and momentum broadening due to finite temperature)
and quantum depletion beyond the mean-field approximation are beyond the scope of our mean-field simulations.

\section{Experimental Results}
\label{expt}

\subsection{Two- and three- site lattices}

The simplest system we perform spectroscopy on is a two-site lattice with tunneling
strength $t = h\times\SI{1598\pm 10}{\hertz}$ and injection tunneling strength
$t_\textrm{inj} = h\times\SI{40\pm 10}{\hertz}$. The eigenenergies for this lattice are simply
$E = \pm t$, and the eigenstates are equal symmetric and anti-symmetric superpositions of atoms at the
left and right site. When we perform spectroscopy on the two-site lattice, we see that
away from $E_\textrm{inj}/t = \pm 1$ the experimental loss time is approximately
$\SI{40}{\milli\second}$. We observe two features of decreased loss times in 
Fig.~\pref{fig:sims}{a}, relating to dips in the data near the expected resonances at $E_\textrm{inj}/t = \pm 1$.
This observation can also be understood as Autler-Townes splitting of the bare probe-injection site transition due to hybidization of the states $p = 2\hbar k$ and $4 \hbar k$ by the applied Bragg field. 

The next simplest system is the three-site lattice with uniform tunneling strength
$t = h\times \SI{2040\pm 3}{\hertz}$ and injection tunneling strength
$t_\textrm{inj} = h\times \SI{50 \pm 3}{\hertz}$.
The energies for this type of lattice are $E =
0, \pm \sqrt{2}t$. When we perform spectroscopy on the three-site lattice, we see a
similar trend as before. When $E_\textrm{inj}$ is nearly resonant with an eigenenergy of the
lattice, the loss time decreases. When $E_\textrm{inj}$ is away from the eigenenergies of
the lattice, the loss time is roughly constant.
One feature of the three-site loss spectrum is that the loss time dips appear to be shifted downward in energy relative to their naive expectation values. Indeed, as seen also in the numerically simulated curve, one should expect a slight downward shift in energy due to the ac Stark shifts of the momentum states, \textit{i.e.}, due to the fact that the strong tunneling links in the lattice induce momentum-dependent light shifts that shift the lattice
site energies relative to the probe site.

For these two simplest cases, we note that the predicted loss curves curves generated from the GPE simulations match
the locations of the dips in the spectrum but do not perfectly match the scale of the
experimentally measured loss time. For the two-site case, the loss time away from
the eigenergies is measured to be approximately $\SI{40}{\milli\second}$, while the
simulation predicts the loss time should be closer to $\SI{90}{\milli\second}$. It is again interesting to note that, as seen more clearly in the simulated spectrum, there is a slight downward shift in energy due to the large system tunneling, which results in momentum-dependent ac Stark shift to the synthetic lattice site energies.

For the
three-site lattice, the loss time away from the eigenergies is measured to be
approximately $\SI{15}{\milli\second}$, while the simulation predicts it should be roughly
$\SI{80}{\milli\second}$. While the GPE simulation does account for the fact that the
cloud separates spatially, thus limiting the coherence of the time evolution, it does not
account for additional loss mechanisms. Two possible mechanisms could include momentum-changing $s$-wave collisions between momentum orders, which scatter atoms into modes outside of those considered, as well as scattering between
thermal and condensed atoms. Furthermore, there are oscillations in the simulated
spectra which are not captured in the experimental data, which is likely due to the fact
that there are additional loss mechanisms and inhomogeneous density shifts that wash them
out.

\begin{figure*}[t]
    \centering
    \includegraphics[width=1.7\columnwidth]{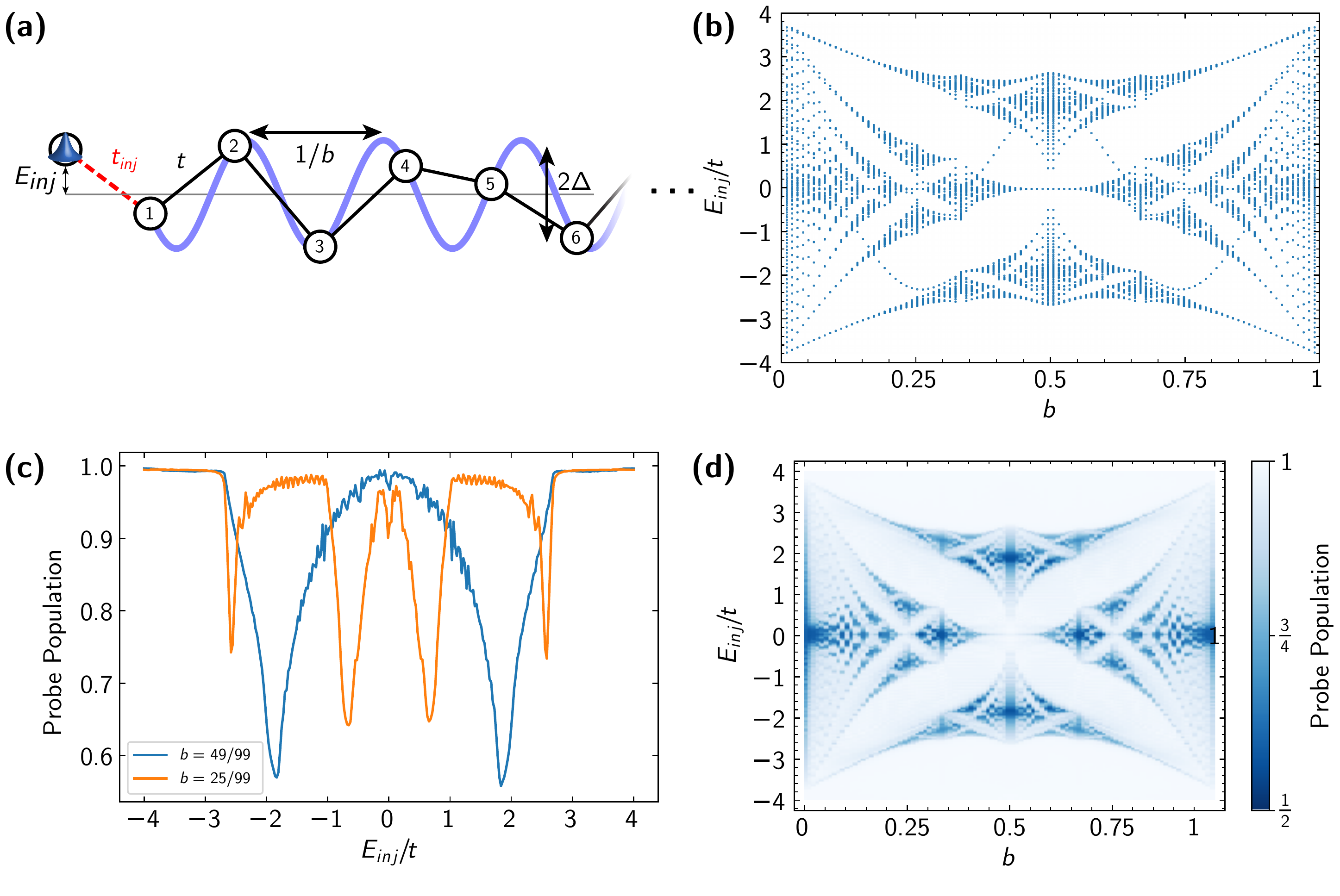}
    \caption{
    \textbf{Injection spectroscopy of an Aubry-Andr\'e-Harper-Hofstadter (AAHH) lattice.}
    \textbf{(a)}~A qualitative depiction of the AAHH lattice under injection spectroscopy. At the left end, a probe site has an adjustable site energy $E_\textrm{inj}$ and is coupled into the lattice with a tunneling rate $t_\textrm{inj}$. In the AAHH lattice, the sites are uniformly coupled with nearest-neighbor tunneling rates $t$, but site-dependent energies $\varepsilon_i$ are quasiperiodically shifted as $\Delta\cos(2\pi b i + \phi)$. Here, $\Delta$ is the modulation depth, $b$ is the incommensurability parameter, and $\phi$ is the phason value.
    \textbf{(b)}~The calculated energy spectrum for the open AAHH chain for $\phi = 0$ and $\Delta/t = 9/5$, with $t=1$, taken for different values of the incommensurability parameter $b$.
    \textbf{(c)}~Two numerical simulations of the injection loss spectrum, relating to the fractional population remaining at the probe site as a function of the injection site energy, for incommensurability parameters $b=49/99$ and $b=25/99$ (both for $\Delta/t = 9/5$ and $t_\textrm{inj}/t = 1/20$). These spectra are $\phi$-averaged over 100 uniformly spaced values of $\phi\in[-\pi,\pi]$. The probe population is calculated after 150 system tunneling times.
    \textbf{(d)}~A composite plot of $\phi$-averaged injection spectra as in (c), but for a larger range of incommensurability parameter values $b\in[0,1]$.
    }
    \label{fig:hof1}
\end{figure*}

We note that the simulations also account for the effects of mode-preserving atomic interactions (effectively local attractive nonlinearities)~\cite{an2018corr}, which effectively serve to shift the energy of the lattice system up relative to the populated probe site. In the two-site case, this upwards shift (relative to the probe) due to interactions and the downwards shift due to the ac Stark effect are nearly perfectly compensated. However, in the three-site case, where we operate with a larger tunneling amplitude, the downwards shift due to the ac Stark effect dominates.

\subsection{Uniform lattice}

The last system we probe experimentally is a uniform chain with tunneling strength
$t=h\times\SI{492\pm 10}{\hertz}$ with the injection link strength
$t_\textrm{inj} = h\times\SI{50\pm 10}{\hertz}$. Such a lattice has eigenenergies ranging from
$-2t$ to $2t$. For this spectrum, instead of measuring the loss rate
by taking measurements at a range of evolution times, we rather directly measured the
amount of population remaining in the $p = 0$ order after an evolution time of
$\SI{20}{\milli\second}$. We then extract the loss time by solving
$P(\SI{20}{\milli\second}) = P(0)e^{-(\SI{20}{\milli\second})/\tau_0}$ for $\tau_0$, assuming a fixed initial population of $P(0) = 6 \times 10^4$ atoms.
%
We see that there is a broad dip in the loss
time for a range of $E_\textrm{inj}$ values near zero energy, with a slight positive shift that is captured well by the simulated curve shown in
Fig.~\pref{fig:sims}{c}.
This slight positive shift is due to the fact that the population in the $p = 0$ order is larger than the
population in any of the other lattice sites, which induces a mean field shift in the site
energy of the $p = 0$ order down by $U\approx 1.5\, t$. This has the effect of shifting the
entire spectrum up in energy by $U$. Note that the interaction shift effect is less
pronounced in the two- and three-site lattices
because those experiments were undertaken with a larger value of the system tunneling $t$. In the uniform lattice, the effect of the light shift is also much
smaller than in our previous two- and three-site lattice cases because the tunneling
strength is roughly a third what we used in the two-site lattice and a fourth of what we used in
the three-site lattice.

\begin{figure*}[t!]
    \centering
    \includegraphics[width=1.9\columnwidth]{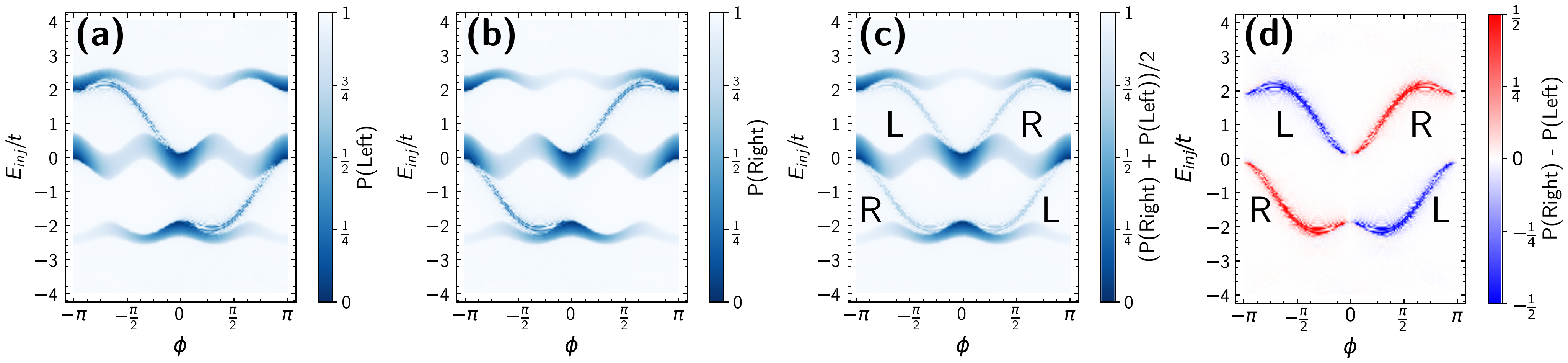}
    \caption{
    \textbf{Boundary-dependent injection into the Aubry-Andr\'e-Harper-Hofstadter (AAHH) lattice.}
    Energy- and phason-resolved injection spectroscopy of a 101-site AAHH lattice lattice for $\Delta/t = 9/5$, an incommensurability ratio $b=1/3$, and $t_\textrm{inj}/t = 1/20$.
    \textbf{(a)}~For probing from the left boundary, fractional population remaining at the probe site (indicated by the color scale at right) as a function of the probe site energy $E_{inj}$ and the phason value $\phi$ after a total of 150 system tunneling times.
    Three bulk energy bands are observed, as well as prominent dispersing inter-band modes.
    \textbf{(b)}~Same measure as in (a), but for the case of probing from the right boundary.
    \textbf{(c)}~The average of the injection spectroscopy signal for the left- and right-sided injection.
    \textbf{(d)}~The difference between the left- and right-sided injection signals, revealing on the presence of the topological boundary states on the right (in red) and left (in blue) system boundaries.
    }
    \label{fig:hof2}
\end{figure*}

\section{Numerical Results}
\label{sims}

\subsection{Aubry-Andr\'e-Harper-Hofstadter model}

The Harper-Hofstadter (HH) model is used to describe the motion of an electron in a lattice
that is placed in a uniform external magnetic field. 
In cold atoms, the two-dimensional HH model been realized for cold atoms using techniques of laser-assisted hopping in real-space lattices~\cite{Aidelsburger2013,Miyake2013}.
The two- and three-dimensional
versions of the HH model can also be effectively reduced to the problem of a one-dimensional periodically modulated lattice model, where the periodicity of the modulation is set by the ratio $\Phi/\Phi_0$, with $\Phi$ the amount of magnetic flux
through a plaquette and $\Phi_0 = h/e$ is the quantum of flux~\cite{Harper1955,
Hofstadter1976}. 
This dimensionally-reduced model, the Aubry-Andr\'e-Harper-Hofstadter (AAHH) model, has played an important role in explorations of localization phenomena with cold atoms~\cite{roati2008anderson,das2019realizing,Luschen-ME,an2021interactions} and for studies of topological edge states  ~\cite{lau2015topological,aidelsburger2015measuring,ni2019observation,ye2022hofstadter}.

One longstanding goal of studies of the Harper-Hofstadter model is to directly measure its fractal energy spectrum, namely the famous ``Hofstadter butterfly.'' While there
are proposals to observe the butterfly spectrum in driven optical lattices~\cite{Rajagopal-spec-quasicrystal}, there has not yet been a method proposed to observe
it in a synthetic lattice. Here, we show that the injection spectroscopy technique may prove useful for the measurement of the HH butterfly spectrum, and for the measurement of the corresponding topological edge states.

To be concrete, the model we consider is given by the AAHH tight-binding
Hamiltonian: 
\begin{align}
\hat{H} = -\sum\limits_{i=1}^{N-1} t(\ket{i}\bra{i+1} + \textrm{h.c.})
+ \sum\limits_{i=1}^N \Delta\cos(2\pi b i + \phi)\ket{i}\bra{i}.
\end{align}
Here, $t$ is the tunneling strength between nearest neighbors, $\Delta$ is the strength of the potential energy modulation, $1/b$ is the periodicity of
the site-energy modulation, and $\phi$ is the phason degree of freedom, relating to a phase shift to the sinusoidal potential modulation. Note that
for rational values of $b$ the site potential modulation is periodic, but for irrational $b$ the site potential energy shifts
will never repeat. This model is shown schematically in Fig.~\pref{fig:hof1}{a}, along with its fractal energy eigenstate spectrum as a function of the incommensurability parameter $b$, shown in Fig.~\pref{fig:hof1}{b}.

In our theoretical study, we first consider the AAHH model with $\Delta/t =
9/5$ for various $b$. The number of sites is $N = 101$. We begin by having all the
population start in the site $\ket{p = 0\hbar k}$, and we let the system evolve for $150$
tunneling times. At the end of each simulation, we record the population that is left in
the $\ket{p = 0\hbar k}$ site. We repeat this calculation for many different
energy offset values of the probe site to get the spectrum for one value of $b$.
Two such loss spectra, for values of $b=24/99 \  (\sim 1/4)$ and $b=49/99 \ (\sim 1/2)$, are shown in Fig.~\pref{fig:hof1}{c}. These spectra reveal four and two primary loss features, respectively, relating to the existence of a corresponding number of bulk mini-bands for these values of the incommensurability parameter $b$.
We can repeat this simulation
for a large range of $b$ values, keeping $\Delta/t$ constant, and we find the
emergence of the famous Hofstadter butterfly spectrum shown in Fig.~\pref{fig:hof1}{d}.
These simulations of tunneling-based loss spectra match qualitatively with the full numerically calculated spectrum shown in
Fig.~\pref{fig:hof1}{b}. Note that some eigenvalues are not represented well in our simulated
spectrum, likely due to the fact that they correspond to eigenvectors that have very
little (or no) weight at the site of injection ($\ket{1}$). We note that these simulations assume a large timescale of relevant tunneling times, however these structures may still be resolved for shorter probing times, especially in the somewhat trivial strong $\Delta$ limit (if averaging over the phason degree of freedom).

In Fig.~\ref{fig:hof2}, we show how one may probe the topological edge states associated with the AAHH spectrum. We fix $b$ to $1/3$ and $\Delta/t$ to $9/5$ and we vary
the value of $\phi$, which is related to the $k_y$ wave vector of the higher-dimensional HH model. We perform two sets of simulations corresponding to injection at opposite sides of the open-boundary AAHH lattice. In one set of simulations, we start in the $\ket{p=0}$ state and
construct the lattice with sites $\ket{p = 2n\hbar k}$, where $n \geq 1$. We call this the
``left injection'' configuration. For a fixed $\phi$, we repeat this simulation for many
different $E_\textrm{inj}$ values in order to produce a loss spectrum. In the alternate set of simulations, we begin in the $\ket{p=0}$ state and we construct the lattice with sites $\ket{p = 2n\hbar k}$, where $n \leq -1$; we call this
the ``right injection'' configuration. The left and right injection spectra are shown in
Fig.~\pref{fig:hof2}{a} and Fig.~\pref{fig:hof2}{b}, respectively. In both spectra, there are three
bands, as well as two modes that disperse between the bulk bands. The combination (average) of the two spectra reveals
the entire spectrum, including bulk and boundary states, as is shown in Fig.~\pref{fig:hof2}{c}. We can also take the difference
between the left and right injection spectra; this subtraction (shown in Fig.~\pref{fig:hof2}{d}) effectively removes the bulk bands, revealing the topological edge modes as well as the edge they live on.

\begin{figure*}[t!]
    \centering
    \includegraphics[width=1.9\columnwidth]{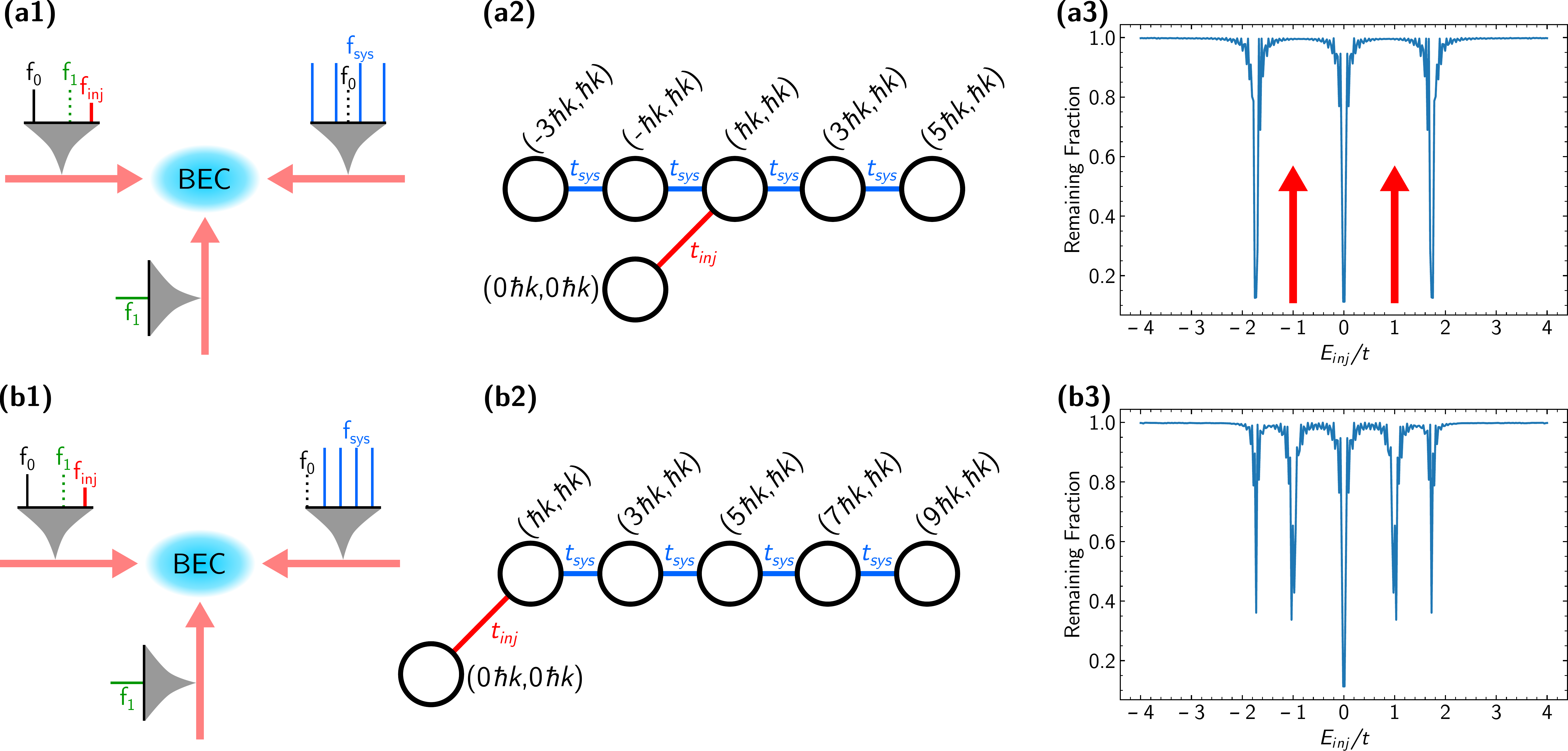}
    \caption{
    \textbf{``Scanning mode'' spectroscopy of momentum state lattices.}
    \textbf{(a1)}~Here, we show the layout for generating a quasi two-dimensional lattice. There
    are two counter-propagating laser beams that generate the lattice, and one beam
    orthogonal to both counter-propagating beams. The orthogonal beam provides the
    link between the injection site and the lattice.
    \textbf{(a2)}~The three beams in (a1) 
    couple different momentum sites as shown here. The transition frequencies in the
    counter-propagating beam are chosen such that the probe site is connected to
    a site in the middle of the lattice.
    \textbf{(a3)}~A simulated loss spectrum for the situation
    depicted in (a2), where initially all of the population is in the injection site. 
    Here, $t_\textrm{inj}/t_\textrm{sys} = 1/20$ and we plot the
    remaining fraction of the population in the probe after 150 tunneling times.
    \textbf{(b1)}~Here we depict the frequencies necessary to make a lattice where the probe
    site is attached to the left edge of a uniform five-site lattice.
    \textbf{(b2)}~The lattice
    beams interfere to make the effective five-site tight binding lattice with uniform
    tunneling, plus a probe site attached to the left edge of the lattice.
    \textbf{(b3)}~Simulated spectrum for the situation depicted in (b2). Again, all of the population
    initially starts in the probe site, and fixing $t_\textrm{inj}/t_\textrm{sys} = 1/20$ we plot the
    remaining probe fraction after 150 tunneling times. Notice
    that there are two additional dips that were not present if we only probed into the center
    of the lattice.}
    \label{fig:scanning}
\end{figure*}

\subsection{Scanning mode spectroscopy}

Here, we describe an extension of the one-dimensional techniques in order to probe the
bulk as well as edge states. As we stated in previous sections, it is possible to miss
certain eigenenergies if the corresponding eigenvector has no weight in the site to
which the probe site is connected to. In order to counteract this, it would be useful
to have a scheme where we can attach the probe site to any site in the lattice.

In order to probe any site in the lattice, in more direct analogy to scanning probe techniques~\cite{STM-RMP}, we can extend the considered one-dimensional momentum-
state lattice to a quasi two-dimensional geometry. In practice, this could be enabled by adding an additional laser beam that propagates in the direction perpendicular two the original two counter-propagating beams, as is shown in
Fig.~\pref{fig:scanning}{a1}. Interference between the frequency component $f_1$ on the up-traveling beam and the frequency component $f_\text{inj}$ on the right-traveling
beam can drive Bragg transitions from the $\ket{0\hbar k, 0\hbar k}$ state to the
$\ket{\hbar k,\hbar k}$ state. We can then construct the frequency tones, $f_\text{sys}$ on the
left-traveling beam such that their interference with the frequency tone $f_0$ on the right-traveling beam resonantly drives transitions between the momentum states $\ket{\hbar k, (2n+1)\hbar k}$
and $\ket{\hbar k, (2n + 3)\hbar k}$, where $n$ is an integer, in order to construct
a one-dimensional lattice of initially unoccupied states, as depicted in Fig.~\pref{fig:scanning}{a2}.
In order to prevent unwanted interference effects, the frequency difference between $f_0$ and $f_1$ are set to be many orders
of magnitude above, say, the recoil energy $E_R = \hbar^2k^2/2M$, while the frequency difference
between $(f_0,f_\text{sys})$ and $(f_1,f_\text{inj})$ are, respectively, on the order of the recoil energy.
As one simple example, we consider probing a uniform one dimensional lattice with some tunneling strength
$t_\textrm{sys}$, by injecting into the center site as shown in Fig.~\pref{fig:scanning}{a2}. When the energy of the probe is scanned, it reveals three loss features, as seen in the simulated
spectrum shown of Fig.~\pref{fig:scanning}{a3}. In this case, two eigenmodes of the system under consideration (the five-site uniform lattice) are not revealed by the loss measurement due to symmetry -- they have exactly zero weight at the site of injection.

In this ``scanning mode'' injection spectroscopy, the frequency spectrum of the left-traveling beam can be altered to that depicted in Fig.~\pref{fig:scanning}{b1}, such that the injection site ($\ket{\hbar k,\hbar k}$) actually resides at the left end of the five-site lattice, as shown in Fig.~\pref{fig:scanning}{b2}. 
The corresponding loss spectrum for this configuration reveals five loss features, corresponding to the full set of eigenstates of this system, including the two modes at energies of $\pm t$ that were missed by the center-site injection.

This simple example shows how tunneling spectroscopy in synthetic lattices may be further extended to a ``scanning mode'' to allow for greater utility in characterizing different model systems. Beyond this, injection spectroscopy in synthetic lattice systems can even be extended to include simultaneous injection at multiple sites of a system (with control of the relative amplitude and phase of injection at different locations). This very unique capability can, for example, be used to perform (approximations to) wavevector-resolved spectroscopy as well as band-specific spectroscopy in multi-band systems with multi-site unit cells.

\section{Conclusion}
\label{conc}

Synthetic lattices in cold atom systems offer a powerful window into the physics of many
condensed matter phenomenon such as localization, topological insulators, and the quantum
Hall effect. In this work, we have shown how it is possible to probe the spectrum of
one-dimensional synthetic lattices made by laser-coupled momentum states. Experimentally, for the examples of two-site, three-site, and many-site uniform lattices, we have shown that it is possible to reproduce the eigenenergy locations through observed loss resonances from an injected probe site. By comparing to numerical (GPE model) simulations, we found that the experimental data exhibited qualitative agreement with the expected locations of resonant loss features in these systems, including mean-field shifts due to atomic interactions.
Additionally, through numerical simulations, we have shown theoretically that this injection spectroscopy technique
could allow for energy spectrum studies of the Aubry-Andr\'e-Harper-Hofstadter model,
opening the way to the measurement of fractal energy spectra and topological boundary states in synthetic lattices.

\section{Acknowledgements}
This material is based upon work supported by the Air Force Office of Scientific Research under Grant No.~FA9550-21-1-0246.
We thank Barry Bradlyn and Jackson Ang'ong'a for helpful discussions.

\bibliographystyle{apsrev4-1}
\bibliography{STM}
\end{document}